\documentclass[%
    reprint,
    nofootinbib,
    nobibnotes,
    amsmath,amssymb,
    aps,
    prl,
    floatfix,
]{revtex4-2}

\usepackage{graphicx}
\usepackage{dcolumn}
\usepackage{bm}
\usepackage{hyperref}

\usepackage[T1]{fontenc}

\usepackage{color}

\begin{document}

\preprint{APS/123-QED}

\title{Quantum entanglement of photon pairs at proton-proton colliders}

\author{Yue Pan}
\email{y.pan@cern.ch}
\affiliation{School of Physics and State Key Laboratory of Nuclear Physics and Technology, Peking University, Beijing, 100871, China}

\author{Yipin Wang}
\email{yipin.wang@cern.ch}
\affiliation{School of Physics and State Key Laboratory of Nuclear Physics and Technology, Peking University, Beijing, 100871, China}

\author{Leyun Gao}
\email{leyun.gao@cern.ch}
\affiliation{School of Physics and State Key Laboratory of Nuclear Physics and Technology, Peking University, Beijing, 100871, China}

\author{Qiang Li}
\email{qiang.li@cern.ch}
\affiliation{School of Physics and State Key Laboratory of Nuclear Physics and Technology, Peking University, Beijing, 100871, China}

\author{Chen Zhou}
\email{chen.zhou@cern.ch}
\affiliation{School of Physics and State Key Laboratory of Nuclear Physics and Technology, Peking University, Beijing, 100871, China}


\begin{abstract}
Diphoton systems, with photon polarizations measurable at low energies, serve as ideal qubits and were first used to demonstrate quantum entanglement. However, due to the current absence of dedicated polarimeters at high-energy colliders, the entanglement properties of diphoton systems remain largely unexplored at the high-energy frontier. Studying quantum entanglement at the high-energy frontier, where particle colliders serve as a natural relativistic laboratory, helps us better understand the quantum nature and seek new physics. In this letter, we propose a novel method to measure the entanglement of diphoton systems at proton-proton colliders. The photon spin analyzing power arises from the Bethe–Heitler process occurring in the tracker, where photons scatter off nuclei to produce electron–positron pairs whose joint angular distribution encodes the polarization of the diphoton system. Simulation results show that, under HL-LHC conditions, the statistical significance of quantum entanglement in the \(\mathrm{Higgs} \to \gamma\gamma\) process is \(0.007\,\sigma\), while measuring the continuum diphoton process \(q\bar{q} \to \gamma\gamma\) alone can reach about \(1.5\,\sigma\).
\end{abstract}

\maketitle


Quantum entanglement is a fundamental feature of quantum systems. Physical systems that have once interacted can remain correlated even after they are separated by large distances and no longer interact with each other. Initially, quantum entanglement and the nonlocality of quantum systems were regarded merely as theoretical debates. It was not until the formulation of Bell’s inequality in 1964 that these concepts became experimentally testable~\cite{Bell:1964kc}. In 1972, 1982, and 1997, the experiments performed by John Clauser, Alain Aspect, and Anton Zeilinger progressively established violations of Bell’s inequality using entangled photon pairs, thereby confirming the existence of quantum entanglement~\cite{Freedman:1972zza,Aspect:1982fx,Bouwmeester:1997slj}. Their pioneering contributions were recognized with the 2022 Nobel Prize in Physics. To date, most experimental studies of quantum entanglement have been carried out in the fields of quantum optics and atomic physics, while measurements at high-energy colliders remain relatively limited. The most prominent example is the measurement of top-quark pair entanglement reported by the ATLAS and CMS Collaborations in 2024~\cite{ATLAS:2023fsd,CMS:2024pts,CMS:2024zkc}. More recently, the ATLAS Collaboration reported measurements of entanglement between Z boson pairs originating from Higgs boson decays~\cite{ATLAS:2026hye}. 
In addition, using helicity-amplitude measurements from the LHCb Collaboration, violations of Bell's inequality in B-meson decays have been established~\cite{Fabbrichesi:2023idl}, and the BESIII Collaboration has observed Bell inequality violations in entangled $\Lambda\bar{\Lambda}$ hyperon pairs~\cite{BESIII:2025vsr}. 

Although photon pairs, as one of the simplest two-qubit systems, have long served as canonical platforms for studying quantum entanglement, no direct measurement of entangled photon pairs has yet been realized at a high-energy collider. The Large Hadron Collider (LHC) at CERN is currently the highest-energy particle collider in operation and provides a unique laboratory for studying quantum phenomena in relativistic and high-energy environments. Its large Higgs-boson data samples collected by the ATLAS and CMS experiments make it possible to investigate quantum entanglement in processes that are inaccessible in traditional quantum-optics experiments. While entangled diphoton systems have been extensively studied in quantum-optics experiments, collider environments offer the opportunity to probe the same quantum correlations in a fundamentally different kinematic regime characterized by relativistic particles and multi-TeV collision energies. Such studies may further deepen our understanding of the nature of quantum entanglement. This study also helps us examine the spin correlation structure of the diphoton system from Higgs decays. Meanwhile, the quantum correlations between diphoton produced in Higgs boson decays are also sensitive to the CP structure of the Higgs sector. The distribution of entanglement observables can also directly reflect the CP structure of the Higgs couplings~\cite{Fabbrichesi:2022ovb,Bishara:2013vya}. 

As the only fundamental scalar particle in the Standard Model, the Higgs boson carries spin zero, which imposes strong constraints on the angular momentum structure of its decay products~\cite{CMS:2014nkk}. In the diphoton decay of the Higgs boson (\(H\to\gamma\gamma\)), the polarization states of the two photons must satisfy both total angular momentum conservation and parity symmetry, resulting in a quantum superposition state with a well-defined structure. Equivalently, only specific helicity combinations are allowed at the amplitude level, leading naturally to entanglement in the polarization space. The effective Lagrangian for the \(H\to\gamma\gamma\) interaction can be written as
\begin{equation}
\mathcal{L}_{\rm eff}\supset
-\frac{1}{4}h(\kappa_\gamma F^{\mu\nu}F_{\mu\nu}
+
\tilde{\kappa}_\gamma F^{\mu\nu}\tilde{F}_{\mu\nu})
\label{eq:Lagrangian},
\end{equation} 
where \(h\) denotes the Higgs field, \(F_{\mu\nu}\) is the electromagnetic
field-strength tensor, \(\tilde F_{\mu\nu}
=\frac12\epsilon_{\mu\nu\rho\sigma}F^{\rho\sigma}\) is its dual tensor. \(\kappa_\gamma\) and \(\tilde{\kappa}_\gamma\) are the coefficients of the CP-even and CP-odd terms. 

Taking into account angular momentum conservation together with the spin-zero nature of the Higgs boson, the general polarization density matrix of the photon pair can be written as
\begin{equation}
\begin{pmatrix}
\sin^2\omega   & 0 & 0 & \sin\omega\cos\omega \mathrm{e}^{-i\delta} \\
 0 & 0 & 0 & 0\\
 0 & 0 & 0 &0 \\
 \sin\omega\cos\omega \mathrm{e}^{i\delta} & 0 &0  &\cos^2\omega 
\end{pmatrix},
\end{equation}
where \(\sin\omega\) and \(\cos\omega\) are the amplitudes of \(|++\rangle\) and \(|--\rangle\) (\(\omega\in[-\pi,\pi]\)) and \(\delta\) is the phase difference between these two states. Whether the state is separable can be determined using the Peres–Horodecki criterion~\cite{Peres:1996dw,HORODECKI19961}. The criterion states that a bipartite system is entangled if and only if there exists a positive map \(\Lambda\) acting on subsystem \(S_A\) such that the density matrix \(\rho\) loses its positivity under the action of \(\Lambda_A\otimes 1_B\),
\begin{equation}
(\Lambda_A\otimes1_B)\rho \not\ge 0 .
\end{equation}
One of the simplest positive maps is the transpose operation. Therefore, if the system is no longer positive semidefinite under partial transposition, the state is entangled. For two-qubit systems, the Peres–Horodecki criterion provides a necessary and sufficient condition for entanglement. Performing the partial transpose of the above polarization density matrix yields the eigenvalues \(\sin^2\omega\), \(\cos^2\omega\), \(\pm\left | \sin\omega\cos\omega \right | \). Therefore, the condition
\(
\left|\sin\omega\cos\omega\right|\neq0
\) is the necessary and sufficient condition for the diphoton quantum state produced in Higgs boson decay to be entangled. For a two-qubit system, the auxiliary matrix is defined as
\(
R=\rho(\sigma_y\otimes\sigma_y)\rho^*(\sigma_y\otimes\sigma_y),
\) where \(\sigma_y\) denotes the Pauli matrix. Denoting by \(r_i\) the square roots of the eigenvalues of \(R\), the concurrence is defined as
\begin{equation}
\mathcal{C}[\rho]=\max(0,r_1-r_2-r_3-r_4)
\end{equation}
where \(r_1\) is the largest eigenvalue square root~\cite{Bennett:1996gf,Wootters:1997id}. Substituting the diphoton density matrix into the above expression, the concurrence is obtained as
\(\left |  \sin2\omega\right | \).

\begin{figure}[htb]
  \centering
  \includegraphics[width=9cm]{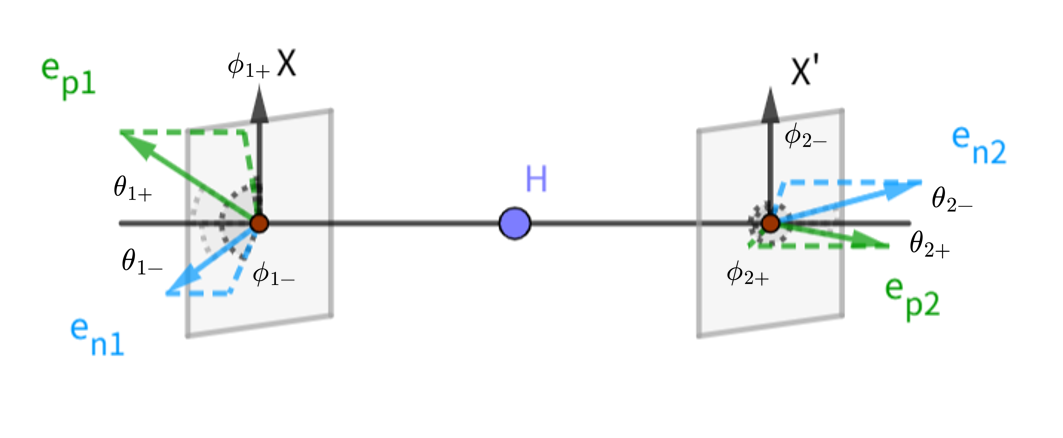}
  \caption{Illustration of the Higgs boson decaying into two photons, with each photon then converting to an electron (labeled as \(\mathrm{e_{n1}}\) and \(\mathrm{e_{n2}}\)) and a positron (labeled as \(\mathrm{e_{p1}}\) and \(\mathrm{e_{p2}}\)). The angles \(\theta_{i\pm}\) and \(\phi_{i\pm}\) are the polar and azimuthal angles of the electron or positron relative to the parent photon's direction, with the reference axis for \(\phi_{i\pm}\) (x-axis) chosen as an arbitrary direction perpendicular to the photon momentum.}
  \label{fig:define}
\end{figure}

Therefore, once the polarization density matrix is reconstructed, one can directly determine whether the quantum state is entangled. Since particle polarization cannot be measured directly by detectors, previously studied bipartite entanglement systems at colliders, such as \(t\bar t\), \(WW\), and \(ZZ\), rely on the angular distributions of the final-state leptons from particle decays to probe polarization information. However, photons at colliders do not decay, rendering this method inapplicable. For high-energy photons, the dominant interaction process in matter is pair production (i.e., the Bethe-Heitler process, hereafter abbreviated as BH), whose cross section is nearly two orders of magnitude larger than that of Compton scattering for \(E_\gamma\ge20\mathrm{GeV}\)~\cite{berger2010xcom}. For polarized photons, the Bethe–Heitler cross section depends on the azimuthal angle \(\phi_\pm\) of the final-state electron~\cite{Bethe:1934za,PhysRev.78.623}. Consequently, pair production provides a natural probe of photon polarization and has already been exploited in astrophysical \(\gamma\)-ray detection experiments~\cite{Barbiellini:2004hf,Depaola:2009zz}. In collider detectors, converted photons correspond precisely to photons undergoing pair production, and the reconstruction of converted-photon information has become a mature component of photon identification techniques~\cite{ATLAS:2018fzd,CMS:2015myp}. Therefore, the polarization density matrix of photons can be reconstructed from the angular distributions of the final-state electrons originating from photon conversions. Analogous to the decay spin-density matrix formalism, we introduce the Bethe–Heitler spin-density matrix to describe the scattering of photons off atoms in the inner detector into electron–positron pairs. It is defined in the Higgs boson rest frame as
\begin{equation}
\tilde{T} = 
\begin{pmatrix}
\sum\limits_{i,j} \big| M^{+}_{ij} \big|^2 & \sum\limits_{i,j} M^{+}_{ij} \big( M^{-}_{ij} \big)^* \\[10pt]
\sum\limits_{i,j} M^{-}_{ij} \big( M^{+}_{ij} \big)^* & \sum\limits_{i,j} \big| M^{-}_{ij} \big|^2
\end{pmatrix},
\end{equation}
where \(M^\pm_{ij}\) are the Bethe--Heitler helicity amplitudes, with \(\pm\) denoting the photon helicity and \(i,j=1,2\) denoting the spin states of the final-state electron and positron. The BH helicity amplitudes can be calculated using the spinor-helicity formalism. To simplify the nuclear form factor, we assume the target atom is at rest in the Higgs boson rest frame. Retaining only the leading-order contribution and choosing an appropriate coordinate system, the amplitude then takes a simplified form~\cite{Bishara:2013vya}:
\begin{equation}
\begin{gathered}
M^-_{11}=-(M^+_{22})^*\simeq\mathcal{G}(q^2)\frac{2\sqrt{2\gamma_+\gamma_-}}{q^2}\left(\frac{1}{1+\gamma^2_+\theta^2_+}-\frac{1}{1+\gamma^2_-\theta^2_-}\right),\\[6pt]
\begin{aligned}
M^-_{\substack{12 \\ 21}}=+(M^+_{\substack{21 \\ 12}})^*
&\simeq\pm\mathcal{G}(q^2)\frac{2\sqrt{2\gamma_+\gamma_-}}{q^2}\frac{\gamma_\mp}{\gamma_++\gamma_-} \\
&\quad\times\left(\frac{\gamma_+\theta_+\mathrm{e}^{-i\phi_+}}{1+\gamma^2_+\theta^2_+}
+\frac{\gamma_-\theta_-\mathrm{e}^{-i\phi_-}}{1+\gamma_-^2\theta^2_-}\right),
\end{aligned}\\[6pt]
M^-_{22}=-(M_{11}^+)^*\simeq0.
\end{gathered}
\end{equation}
Here \(\gamma_\pm=E_\pm/m_e\) are the Lorentz factors of the electron and positron, \(q\) is the momentum transfer to the nucleus, and \(\mathcal{G}(q^2)\) is the nuclear form factor. The kinematic variables of the final-state particles are illustrated in Fig.~\ref{fig:define}. Here, \(i=1,2\) labels the two photons, while \(+\) and \(-\) denote the positron and electron, respectively. The polar angle \(\theta_{i\pm}\) is defined as the angle between the lepton momentum and the momentum of the corresponding photon. The azimuthal angle \(\phi_{i\pm}\) is defined as the angle between the projection of the lepton momentum onto the photon polarization plane and the x-axis, where the x-axis is perpendicular to the photon momentum direction and lies within the photon polarization plane.

The squared amplitude can therefore be expressed as
\begin{equation}
\left |\mathcal{M}  \right |^2\simeq\mathrm{Tr}[\rho(\tilde{T}_1\otimes \tilde{T}_2)].
\end{equation}
After simplification, one obtains
\begin{equation}
\begin{split}
\left|\mathcal{M}\right|^2 \propto 
&\sum_{1_{ij}}\left|M_{1_{ij}}^-\right|^2
\sum_{2_{ij}}\left|M_{2_{ij}}^-\right|^2 \\
&+ 4\sin2\omega\,
\mathrm{Re}\!\left[\mathrm{e}^{-i\delta}
M^-_{1_{12}}M^-_{1_{21}}
M^-_{2_{12}}M^-_{2_{21}}\right]
\end{split},
\end{equation}
where \(M^\pm_{1_{ij}} \) is the helicity amplitude of the first photon and \(M^\pm_{2_{ij}}\) is the second's. 
The differential cross section can be written as
\begin{equation}
\mathrm{d}\sigma\propto\left|\mathcal{M}\right|^2\prod_{\substack{i=1,2 \\ \lambda=\pm}}\left[\left|\mathbf{p}_{i\lambda}\right|\mathrm{d}\Omega_{i\lambda}\right]\mathrm{d}\Omega_1\mathrm{d}E_{1-}\mathrm{d}E_{2-}.
\end{equation}
Since the photon polarization is encoded in the angular distributions of the final-state electrons, the observable sensitive to the entanglement of the photon polarization state should depend on the azimuthal angles \(\phi_{i\pm}\) of the electron and positron. However, the momentum transfer \(q^2\) appearing in the amplitude contains the combination \(\phi_{i+}-\phi_{i-}\), implying that the variables \(\phi_{i\pm}\) are not separable from the remaining degrees of freedom \(\theta_{i\pm}\) and \(E_{i-}\).

To disentangle these variables, one may perform a change of variables for the four azimuthal angles \(\phi_{i\pm}\) (\(i=1,2\)), replacing them by \(\phi_{i+}-\phi_{i-}\) and \(\phi_{i+}\) (\(i=1,2\)). In this parametrization, the variables \(\phi_{i+}\) become separable from the remaining degrees of freedom \(\theta_{i\pm}\), \(E_{i-}\), and \(\phi_{i+}-\phi_{i-}\), allowing the latter to be integrated out first. The integration range of \(\phi_{i+}-\phi_{i-}\) is given by \(\phi_{i+}\pm\pi\). Exploiting the \(2\pi\)-periodicity manifest in the amplitudes, the integration interval can be shifted such that its boundaries no longer depend on \(\phi_{i\pm}\). After integrating over the remaining degrees of freedom, the cross section reduces to
\begin{equation}
\begin{split}
&A_{E_{i-},\theta_{i\pm},\phi_{i+}-\phi_{i-}} \\
&\quad + \sin2\omega\,
\mathrm{Re}\Bigl[
e^{-2i(\phi_{1+}+\phi_{2+})-i\delta}
B_{E_{i-},\theta_{i\pm},\phi_{i+}-\phi_{i-}}
\Bigr],
\end{split}
\end{equation}
where \(A_{E_{i-},\theta_{i\pm},\phi_{i+}-\phi_{i-}}\) and \(B_{E_{i-},\theta_{i\pm},\phi_{i+}-\phi_{i-}}\) are expressions involving the other degrees of freedom, which are separable from the observable \(\phi=\phi_{1+}+\phi_{2+}\).
Next, one may redefine the variables \(\phi_{i+}\) in terms of
\(
\phi_{1+}+\phi_{2+}\),
\(\phi_{1+}-\phi_{2+}
\). Using a wrapping procedure, the new integration variables are mapped onto the interval \([-\pi,\pi]\): whenever an angle exceeds \(\pi\), one subtracts \(2\pi\), while for angles smaller than \(-\pi\), one adds \(2\pi\). In this way, the original range \([ -2\pi,2\pi ]\) is folded into \([-\pi,\pi]\). Using
\(
\phi \equiv \phi_{1+}+\phi_{2+}
\)
as the observable, the differential cross section can finally be expressed as
\begin{equation}
\frac{\mathrm{d}\sigma}{\mathrm{d}\phi}
\propto
\mathcal{A}
+
\sin2\omega
\mathcal{B}\cos(2\phi+\delta)
\label{eq:phi_dist},
\end{equation} where \(\mathcal{A}\) and \(\mathcal{B}\) are the coefficients after integrating over the other degrees of freedom.

The Monte Carlo simulation program developed in this work is based on the \( \texttt{G4BetheHeitler5DModel} \) framework~\cite{BERNARD201885}. The squared BH helicity amplitudes derived in the previous section are adopted directly as the differential cross section in the acceptance-rejection sampling procedure. Since the cross-section formula implemented in the original program yields distributions identical to those obtained from the squared amplitudes for all degrees of freedom, the replacement can be performed straightforwardly.We sample the phase space using the five physical degrees of freedom in the laboratory frame: \(
E_-\), \( \theta_\pm\), \( \phi_\pm.
\)

\begin{figure}[htb]
  \centering
  \includegraphics[width=9.5cm]{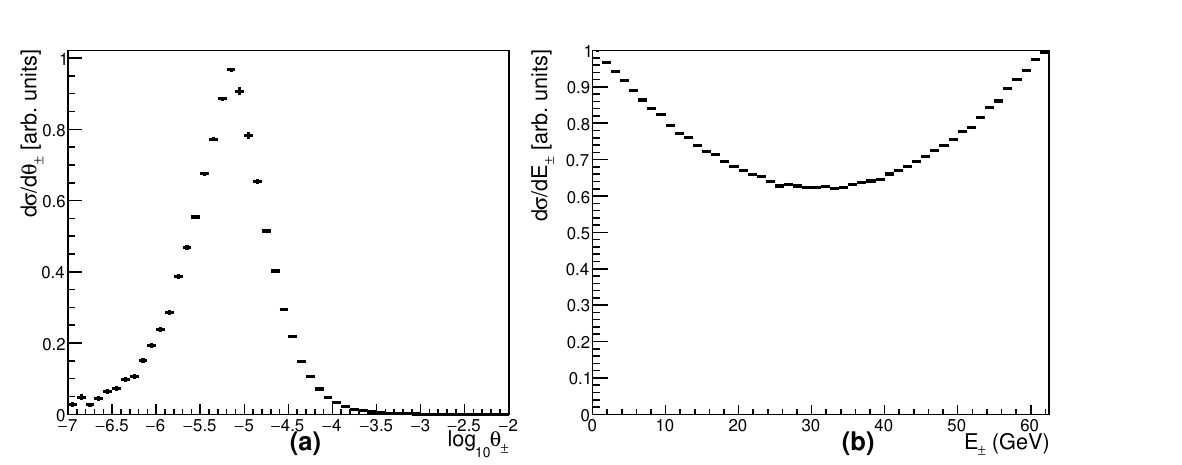}
  \caption{(a) Differential cross-section of the positron polar angle \(\theta\) in the Bethe-Heitler process;
(b) Differential cross-section of the positron energy in the Bethe-Heitler process.
For both (a) and (b), the electron distributions are identical to those of the positron due to symmetry.}
  \label{fig:theta_E}
\end{figure}

The simulated distributions of the kinematic variables for a single-photon BH process are shown in Fig.\ref{fig:theta_E}. Since the squared helicity amplitudes depend only on the combination \(\phi_{i+}-\phi_{i-}\), the individual variables \(\phi_{i\pm}\) are uniformly distributed. These distributions are consistent with the results obtained from direct numerical integration of the amplitude formulas. The numerical integrations are performed using the Cuba library~\cite{Hahn:2004fe}.

Because the quantum correlation of photon polarization originates from the interference between positive- and negative-helicity states, one cannot simply combine two independent BH processes. Likewise, it is insufficient to generate the \(H\to\gamma\gamma\) process using standard event generators such as MadGraph or Sherpa and subsequently feed the final-state photon information into a Geant4 simulation. Directly specifying helicity states of the photons fails to preserve the interference between the diphoton polarization amplitudes. The resulting azimuthal-angle distributions correspond merely to two uncorrelated unpolarized photons. Therefore, retaining the helicity interference requires performing acceptance--rejection sampling based on the squared total amplitude.

The full process from \(H\to\gamma\gamma\) to photon conversion is simulated as follows. The total process's degrees of freedom are sampled uniformly over the phase space, with \(\log_{10}\theta_{i\pm}\in[-7,-2]\), \(\phi_{i\pm}\in[-\pi,\pi]\), and \(E_{i-}\in[m_e,E_{i\gamma}-m_e]\) (where \(E_{i+}=E_{i\gamma}-E_{i-}\)). These sampled variables are then substituted into the total cross section formula, multiplied by the corresponding Jacobian determinant, to compute the final probability density function, which is subsequently used in an acceptance-rejection step to generate the final event sample.

\begin{figure}[htb]
  \centering
  \includegraphics[height=5.3cm]{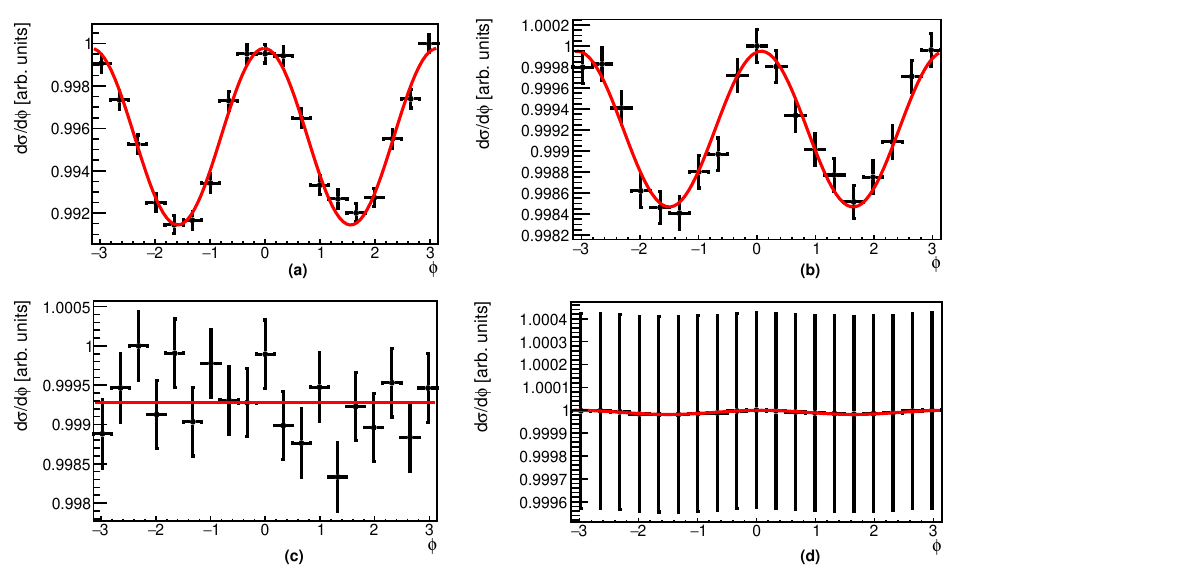}
  \caption{Entanglement measurement phi distribution: (a) Signal process \(H\to\gamma\gamma\) without any cuts;
(b) Signal process \(H\to\gamma\gamma\) with \(\theta_{ll} > 10^{-4}\) cut (\(\theta_{ll}\) is the opening angle between the electron and positron momenta);
(c) Background process \(q\bar{q}\to\gamma\gamma\) (uniform distribution);
(d) Signal + background combined according to the measured ratio (1.2\%) in the signal window. All distributions are normalized. In all panels, the black data points represent the Monte Carlo simulation results, and the red solid curves show the corresponding fitted functions.}
  \label{fig:phi_distribution}
\end{figure}

The resulting distribution of the observable \(\phi\) is shown in Fig.\ref{fig:phi_distribution} (a). With sufficiently large statistics, a clear \(\cos2\phi\) modulation can be observed. The parameters of the normalized distribution function \begin{equation}
1+\alpha\cos(2\phi+\delta)
\end{equation}
are extracted using the maximum-likelihood method, yielding \[\alpha=(4.18 \pm 0.14)\times10^{-3},\qquad\delta=(0.00 \pm 0.03)\mathrm{rad},\] where \(\alpha=\sin2\omega\mathcal{B}/\mathcal{A}\), with \(\mathcal{A}\) and \(\mathcal{B}\) defined in Eq.\eqref{eq:phi_dist}. The phase parameter \(\delta\) corresponds to the relative phase between the \(|++\rangle\) and \(|--\rangle\) helicity amplitudes and is therefore related to the ratio \(\tilde{\kappa}_\gamma/\kappa_\gamma\) in Eq.\eqref{eq:Lagrangian}. The likelihood function employed in the fit is the Poisson likelihood function. The likelihood function employed in the fit is the Poisson likelihood function.

The spatial resolution of the CMS silicon pixel detector is of the order of 10\(\mathrm{\mu m}\), while the innermost pixel layers are located a few centimeters
from the interaction point~\cite{CMS:2008xjf,CMS:2009bxg}. A simple estimate therefore suggests
that opening angles as small as \(\mathcal{O}(10^{-4})\) may be experimentally resolvable. Motivated by this consideration, we impose the requirement \(\theta_{ll}>10^{-4}\) on the opening angle between the electron and positron momenta, denoted by \(\theta_{ll}\). The distribution after requiring
\(
\theta_{ll}>10^{-4}
\)
is shown in Fig.\ref{fig:phi_distribution} (b) Applying the same maximum-likelihood procedure to the corresponding distribution gives \[\alpha=(7.26 \pm 0.51)\times10^{-4},\qquad\delta=(0.00 \pm 0.04)\mathrm{rad}.\] This selection suppresses the amplitude of the \(\cos2\phi\) modulation and consequently reduces the sensitivity.

A large continuum background exists in the diphoton invariant-mass spectrum. The dominant contributions arise from nonresonant diphoton production processes,
\(
q\bar q\to\gamma\gamma\), \(gg\to\gamma\gamma\), \(qg\to\gamma\gamma
q\).
Since the contribution from \(gg\to\gamma\gamma\) is relatively small, being approximately an order of magnitude smaller than the \(q\bar q\to\gamma\gamma\) process within the signal window~\cite{Catani:2018krb}, it is neglected in this work. At leading order, the diphoton polarization density matrix for the \(q\bar q\to\gamma\gamma\) process can be written as
\begin{equation}
\begin{pmatrix}
0 & 0 & 0 & 0\\
0 & D & F & 0\\
0 & F^* & 1-D & 0\\
0 & 0 & 0 & 0
\end{pmatrix},
\end{equation}
where \(D=|\mathcal{M}_{+-}|^2/(|\mathcal{M}_{+-}|^2+|\mathcal{M}_{-+}|^2)\) is the normalized weight of the \(|+-\rangle\) helicity configuration, \(1-D\) is the corresponding weight of the \(|-+\rangle\) configuration, and  \(F\) denotes the interference term between the two helicity amplitudes..

Similar to the analysis of \(H\to\gamma\gamma\), only the \(|+-\rangle\) and \(|-+\rangle\) diphoton helicity states contribute in the \(q\bar q\to\gamma\gamma\) process, while all other helicity amplitudes vanish. The squared total amplitude is therefore given by
\begin{equation}
\begin{split}
|\mathcal{M}|^2
\propto\,
\mathcal{G}_1\mathcal{G}_2
\Biggl(
&\sum_{pq}|\mathcal{M}_{pq}|^2
\sum_{1_{ij}}|M^-_{1_{ij}}|^2
\sum_{2_{ij}}|M^-_{2_{ij}}|^2 \\
&+ 8\mathcal{M}_{+-}\mathcal{M}_{-+}
\,\mathrm{Re}\Bigl[
e^{-i\delta}
M^-_{1_{12}}M^-_{1_{21}}
\bigl(M^-_{2_{12}}M^-_{2_{21}}\bigr)^*
\Bigr]
\Biggr),
\end{split}
\end{equation} where \(\mathcal{G}_i=\mathcal{G}(q_i^2)\) is nuclear form factor of conversion and \(\mathcal{M}_{pq}\) is the tree amplitude of \(q\bar{q}\to\gamma\gamma\), with \(p,q=\pm\) denoting the helicities of the two outgoing photons.

After simplification, the distribution is found to be flat with respect to the observable \(\phi\equiv \phi_{1+}+\phi_{1-}\), while exhibiting a
\begin{equation}
1+\alpha'\cos(2\phi'+\delta')
\end{equation}
dependence with respect to
\(\phi'=\phi_{1+}-\phi_{2+}.
\)
\(\alpha'\) and \(\delta'\) are the coefficients of this distribution.
In principle, the entanglement in the \(q\bar q\to\gamma\gamma\) process may also be probed through the observable \(\phi'\). A flat \(\phi'\) distribution corresponds to the absence of entanglement, whereas the presence of a \(\cos2\phi'\) term indicates entanglement.

The \(q\bar q\to\gamma\gamma\) process is simulated in a similar manner. The \(H\to\gamma\gamma\) decay is isotropic and the Higgs boson invariant mass is fixed. Consequently, the solid angle of the diphoton system can be integrated out directly, while the photon energies are fixed by
\(
\delta(M_h-E_{\gamma1}-E_{\gamma2}).
\)
As a result, the total signal process (\(H\to\gamma\gamma\) with both photons converted) contains only ten degrees of freedom. In contrast, the cross section for \(q\bar q\to\gamma\gamma\) depends explicitly on the scattering angle \(\Theta\) (as noted previously, only the dominant \(q\bar q\to\gamma\gamma\) background is considered, while the smaller \(gg\to\gamma\gamma\) contribution is neglected), and the invariant mass of the quark pair varies within a finite interval corresponding to the signal window. The full process therefore contains twelve degrees of freedom. In the quark-pair center-of-mass frame, the invariant mass of the quark pair equals that of the photon pair. Hence, the diphoton invariant mass discussed below also refers to the invariant mass of the quark pair.

Three additional variables are introduced in the main program to account for the extra degrees of freedom in the \(q\bar q\to\gamma\gamma\) process, namely the longitudinal momentum fractions \(x_1\) and \(x_2\) of the quark and antiquark in the protons and the center-of-mass scattering angle \(\Theta\). Due to the transverse-momentum requirements imposed on photons in experimental analyses, the simulation requires
\(
p_T>20~\mathrm{GeV}.
\)
The allowed range of \(\Theta\) is constrained by both the \(p_T>20\,\mathrm{GeV}\) requirement on the photons and the detector acceptance in pseudorapidity \(\eta\). The random variables \(x_1\) and \(x_2\) represent the momentum fractions carried by the quarks inside the protons, with
\(x_1,x_2\in[0,1]\). The invariant-mass range is chosen as \(M_{\gamma\gamma}\in[120,130]~\mathrm{GeV},\)
and the proton parton distribution functions are obtained using LHAPDF\footnote{ \( \texttt{NNPDF23\_nnlo\_as\_0118\_qed} \) } . Compared with the total signal process process, the output data events for the background simulation additionally record the photon scattering angle \(\Theta\) and the diphoton invariant mass \(M_{\gamma\gamma}\), besides the kinematic variables of the four final-state leptons.

The resulting background distribution in \(\phi\) is shown in Fig.\ref{fig:phi_distribution} (c). A chi-square test yields
\begin{equation}
\chi^2/\mathrm{ndf}=1.03,
\end{equation}
indicating that the background distribution is consistent with a flat distribution. The flat background model is combined with the signal model described by the distribution
\(
1+\alpha\cos(2\phi+\delta).
\)

The cross section for the \(q\bar{q}\to\gamma\gamma\) process is approximately \(31.4~\mathrm{pb}\) at the Run-2 center-of-mass energy of \(13~\mathrm{TeV}\)~\cite{ATLAS:2021mbt}. Based on the fraction of converted photons relative to the total photon yield reported in Table 2 of Ref.~\cite{ATLAS:2018fzd}, the average fraction of converted photons in the \(H\to\gamma\gamma\) process is 0.270. With a cross section of \(67~\mathrm{fb}\) for \(H\to\gamma\gamma\), the expected number of events in which both photons are converted is estimated to be approximately 14650 for the High-Luminosity Large Hadron Collider (HL-LHC) integrated luminosity of \(3000~\mathrm{fb}^{-1}\)~\cite{ATLAS:2022fnp,ATLAS:2023fsd}. For the background process \(q\bar q\to\gamma\gamma\), the number of events within the diphoton invariant mass signal window is approximately 1222110~\cite{ATLAS:2021mbt} requiring both photons to be converted.

The previously generated the signal process Monte Carlo samples and the calculated background model are then rescaled according to their expected event yields at HL-LHC luminosity. The resulting distribution is shown in Fig.\ref{fig:phi_distribution} (d), where the error bars correspond to the Poisson uncertainties in each bin.

The likelihood ratio and the corresponding statistical significance are evaluated, yielding
\(Z=0.007\). If the detector's constraint on \(\theta_{ll}>10^{-4}\) is not considered, the significance will be \(0.04\,\sigma\). The statistical significance is evaluated using the likelihood-ratio test between the null hypothesis of a flat distribution and the signal hypothesis described by \(1+\alpha\cos(2\phi+\delta)\). We also attempted the event categorization strategy similar to that used by CMS or ATLAS in the Higgs discovery, filtering and classifying events such that in certain categories the signal fraction was higher and the significance was improved~\cite{ATLAS:2012yve,CMS:2012qbp}, which can reach approximately \(0.015\,\sigma\).

As discussed above, the quantum entanglement in the \(q\bar{q}\to\gamma\gamma\) process can also be investigated using the same framework. Unlike the study of \(H\to\gamma\gamma\) entanglement, where the diphoton invariant mass must be restricted to the Higgs signal window, no such requirement is necessary when considering the \(q\bar{q}\to\gamma\gamma\) process alone. Therefore, the proton--proton center-of-mass energy is taken to be \(13~\mathrm{TeV}\), the diphoton invariant mass range is chosen as
\(M_{\gamma\gamma}\in[40,500]~\mathrm{GeV}\),
and the photon transverse momentum is required to satisfy
\(p_T>20~\mathrm{GeV}\).
Following the detector acceptance criteria adopted in experimental measurements of \(pp\to\gamma\gamma\) production~\cite{ATLAS:2021mbt}, the photon pseudorapidity is required to satisfy
\(|\eta|<2.37
\) and \(|\eta|\notin[1.37,1.52]
\).

Since the \(q\bar{q}\to\gamma\gamma\) matrix element depends on the scattering angle \(\Theta\), and consequently on the photon pseudorapidity \(|\eta|\), the cross section is integrated over the allowed pseudorapidity regions. Combining this with the photon conversion probability in the corresponding detector regions and integrating over the relevant phase space yields an average conversion probability of approximately 0.382, based on the data in Table 2 of~\cite{ATLAS:2018fzd}. For the HL-LHC integrated luminosity of
\(3000~\mathrm{fb}^{-1}\)~\cite{Apollinari:2017lan},
the expected statistical significance is estimated to be approximately
\(1.5\,\sigma\). If the detector's constraint on \(\theta_{ll}>10^{-4}\) is not considered, the significance will be \(2.1\,\sigma\). This statistical significance is also derived from the likelihood ratio of the flat hypothesis against the \(1+\alpha'\cos(2\phi'+\delta')\) distribution.

Compared with the signal process process, the \(q\bar{q}\to\gamma\gamma\) channel benefits from a much larger event yield, a broader accessible range of \(M_{\gamma\gamma}\), and a negligible background contribution relative to the signal. Consequently, its expected sensitivity is substantially larger. These results suggest that the observation of quantum entanglement in the \(q\bar{q}\to\gamma\gamma\) process at high-energy colliders is considerably more promising and merits further investigation.

In this work, we have explored the feasibility of probing quantum entanglement through the \(H\to\gamma\gamma\) decay channel at high-energy colliders. Based on the diphoton polarization density matrix and the helicity amplitudes of the Bethe–Heitler conversion process, we established a theoretical framework connecting Higgs diphoton decays to the angular distributions of converted electron pairs and derived a \(\cos 2\phi\) observable directly sensitive to quantum entanglement. Combining multidimensional numerical integration with Monte Carlo simulations based on the Geant4 framework, we demonstrated that the diphoton entanglement information can be successfully reconstructed in a pure signal sample. However, after including the dominant continuum background from \(q\bar{q}\to\gamma\gamma\) production under HL-LHC conditions, the statistical significance of the \(H\to\gamma\gamma\) signal is found to be severely suppressed, making a direct observation extremely challenging.

In contrast, the \(q\bar{q}\to\gamma\gamma\) process itself exhibits observable entanglement signatures and offers substantially better experimental prospects at high luminosity. Our results indicate that measurements based on the angular distributions of converted photons provide a novel approach to studying quantum entanglement at high-energy colliders. This work establishes a foundation for future investigations at the HL-LHC and prospective Higgs factories, and highlights a promising interface between high-energy physics and quantum information science.

\begin{acknowledgments}
Acknowledgments: We thank Dr. Hantian Zhang for helpful discussions, which have greatly assisted us in understanding the physics processes in this study and in deriving the cross-section formulas. 
This work is supported by the National Key Research and Development Program of China (Grant No. 2023YFA1605800), the National Natural Science Foundation of China (Grant No. 12522507), and the State Key Laboratory of Nuclear Physics and Technology, Peking University.
\end{acknowledgments}


\nocite{*}

\bibliography{apssamp}

\end{document}